\documentclass{ws-ijmpcs}

\begin{document}

\markboth{Courtoy, Liuti}
{$\alpha_s$ from hadron structure phenomenology.}

\catchline{}{}{}{}{}
%
%%%%%%%%%%%%%%%%%%%%%%%%%%%%%%%%%%%%%%%%%%%%%%%%%%%%%%%%%%%%%%%%%%%%

\title{THE STRONG COUPLING CONSTANT FROM HADRON STRUCTURE PHENOMENOLOGY.}

\author{A.~COURTOY\footnote{
Speaker}}

\address{IFPA, AGO Department, Universit\'e de Li\`ege\\
Liege, Belgium.\\
aurore.courtoy@ulg.ac.be}

\author{SIMONETTA~LIUTI}

\address{{\it Department of Physics, University of Virginia, 382 McCormick Rd. \\
Charlottesville, VA 22904, USA
\\
sl4y@virginia.edu}}

\maketitle

\begin{history}
\received{Day Month Year}
\revised{Day Month Year}
\end{history}

\begin{abstract}
We present recent developments on the role of the running coupling constant at the intersection of perturbative and nonperturbative QCD. 
A number of experiments show a smooth transition from small to large scales given by the four-momentum transfer in the reactions. This is at variance 
with perturbative QCD where the running coupling constant becomes infinite when the scale equals $\Lambda_{\mbox{\tiny QCD}} $. 
Approaches using an effective coupling constant could help interpret the opposite trend of data as compared to standard perturbative QCD predictions. 
We give an overview  of the role of the coupling constant in the procedure to match nonperturbative hadronic model to perturbative QCD and we propose an extraction of an effective coupling constant from inclusive electron proton scattering data at large Bjorken $x$. 
%On the other hand, the Parton Distribution Functions provide a framework to study the transition from nonperturbative model interpretations to perturbative QCD.  
%The strong coupling constant plays a central role in the QCD evolution of parton densities. 
%We extend the standard procedure to match nonperturbative models to perturbative QCD, using experimental data,  with the nonperturbative generalization of the QCD running coupling and use this new development to understand why perturbative treatments are working reasonably well in the context of hadronic models.
%Although these two approaches have been considered so far complementary to each other, a unified description might derive through the definition of the effective coupling, as they both broaden the ways of analyzing the freezing of the running coupling constant.

\end{abstract}

\ccode{PACS numbers: 11.25.Hf, 123.1K}

\section{Hadron Phenomenology}	

The description of the internal structure of the strongly interacting particles is one of the key goals of QCD.  At moderate energy scales, the  hadronic representation supersedes the partonic description, making it challenging to describe the dynamics of scattering processes and hadronic structure. 
On the other hand, it is well known that the Parton Distribution Functions (PDFs) present a framework for connecting the low and high-energy regimes. 
Deep Inelastic processes allow us to look with a good resolution inside the hadron and  to resolve the very short distances, {\it i.e.} small configurations of quarks and gluons. The insight into the structure of hadrons is regulated by factorization theorems:  the large virtuality of the photon, $Q^2$, involved in DIS processes allows for the factorization of their amplitudes into hard (perturbative) and soft (nonperturbative) contributions. At short distances, one singles out a hard scattering process described through Perturbative QCD (PQCD). The large distance part of the process, {\it i.e.} the PDF, reflects 
%how the target reacts to the probe, or 
how the quarks and gluons are distributed inside the target. 

So far, we have introduced two different concepts: a first transition from the hadronic to partonic representation is defined by a {\it  hadronic scale}
%The hadronic scale is found to be 
%$\mu_0^2 \sim 0.1$ GeV$^2$
of a few hundred MeV$^2$ (see {\it e.g.} discussion in Ref.[\refcite{Traini:1997jz}]);  a second scale, namely,  the {\it   factorization scale}, typically $\gtrsim 1$ GeV$^2$, is introduced through the virtuality of the photon. 

Although, as we have just explained, the perturbative stage of a hard collision is distinct from the nonperturbative regime characterizing the hadron structure, experimental observations suggest that, in specific kinematical regimes, both the perturbative and nonperturbative stages arise almost ubiquitously, in the sense that the nonperturbative description follows the perturbative one. This third concept is known as {\it parton-hadron duality}, and we will here understand it as being yet another manifestation of the perturbative to nonperturbative transition in QCD.

\subsection{ Can we extract $\alpha_s$ in the infrared regime from hadronic phenomenology?}

   The standard procedure to fix the hadronic (nonperturbative) scale  pushes perturbative QCD to its limit. The hadronic scale turns out to be of a few hundred MeV$^2$, where the strong coupling constant has already started approaching its Landau pole. 
However, the relative stability of the N$^m$LO evolution 
 %converges very fast, 
 is what justifies the perturbative approach.
   %Consequently, the behavior of the strong coupling constant plays a central role in the QCD evolution of parton densities. 
   An even lower hadronic scale makes sense when considering a freezing of the coupling constant in the infrared region. 
   The standard scheme of scale fixing can then be extended to a nonperturbative evolution framework where the effective coupling is free of Landau pole, {\it e.g.} Refs.~[\refcite{Cornwall:1982zr}--\refcite{Shirkov:1997wi}], while being parameterized with a physical set of parameters~[\refcite{Courtoy:2011mf,Courtoy:2011qa}].
 
On the other hand, the factorization scale fixing procedure 
is carried out entirely within the domain of PQCD, thus relying on the knowledge of $\alpha_s$. Although these two approaches have been considered so far complementary to each other, a unified description might derive through the definition of the effective coupling, as they both broaden the ways of analyzing the freezing of the running coupling constant.
It is in this direction that  the new procedure proposed in Refs.~[\refcite{Courtoy:2011mf,Courtoy:2011qa}] broadens the ways of analyzing the freezing of the running coupling constant: T-odd TMDs are possible candidates to study the behavior of $\alpha_s$ at intermediate and low $Q^2$, along with the observables proposed in Ref.~[\refcite{Deur:2005cf}] (polarized DIS sum rules) and in~[\refcite{Liuti:2011rw}] (perturbative evolution of large $x$ proton structure functions).

In this contribution we focus on the analysis of Ref.~[\refcite{Liuti:2011rw}], where the implications of  parton-hadron duality are explored in the large $x$ region of inclusive electron proton scattering experiments (Bloom--Gilman duality~[\refcite{Bloom:1970xb}]). Our ultimate goal is to provide a procedure on how to match nonperturbative models to PQCD, using experimental data.
The relevant kinematical variables are: $x=Q^2/2M\nu$ ($M$ being the proton mass and $\nu$ 
the energy transfer in the lab system), the four-momentum transfer, $Q^2$, and the invariant mass for the proton, $P$, and virtual photon, $q$, system, $W^2=(P+q)^2$  
($W^2 = Q^2(1/x-1)+M^2$).  For
large values of Bjorken $x  \geq 0.5$, and $Q^2$ in the multi-GeV$^2$ region, one has 
$W^2 \leq 5$ GeV$^2$, {\it i.e.}  the cross section is dominated by resonance formation. While it is impossible to reconstruct the detailed structure of the proton's resonances, these remarkably follow the PQCD predictions when averaged over  $x$.
Although Bloom--Gilman duality was observed at the inception of QCD, quantitative analyses could be attempted only more recently, having at disposal the extensive, high precision data from Jefferson Lab~[\refcite{Melnitchouk:2005zr}]. PQCD-based studies~[\refcite{Liuti:2011rw,Bianchi:2003hi}--\refcite{Liuti:2001qk}], have been presented that include  higher-twist contributions or, more generally, the evidence for nonperturbative inserts, which are required to achieve a fully quantitative fit.
In addition, NLO PQCD evolution at large $x$ can be sensitive to Large $x$ Resummation (LxR) effects. The consequence of LxR is a shift of the scale at which $\alpha_s$ is calculated
to lower values, with increasing $x$ (see for instance  Refs.~[\refcite{Brodsky:1979gy,Pennington:1982kr}]). 
%%
%In other words, the only freedom allowed in a pQCD approach is in the definition of the coupling constant.
This introduces a  model dependence within the PQCD approach in that  the value of the QCD running coupling  in the 
infrared region being regulated by LxR (as we explain below) simultaneously leads to a suppression of higher-twist effects. The higher-twist effects get, in fact, absorbed in the coupling's infrared behavior.

In Section 2  we single out the nonperturbative outcome from a quantitative analysis of the Bloom--Gilman parton-hadron duality. Specifically, we analyze the role of the running coupling constant in the infrared region in tuning the experimental data~[\refcite{Liuti:2011rw,Bianchi:2003hi,our}]. 
The new approach on the freezing of the running coupling constant and its role on the extraction of TMDs from experiment [\refcite{Courtoy:2011mf,Courtoy:2011qa}] is summarized in Section 3. In Section 4 we draw our conclusions.

\section{From perturbative to nonpertubative QCD}

Bloom--Gilman duality implies a one-to-one correspondence between the behavior of the structure function, $F_2$, for unpolarized electron proton scattering in the resonance region, and in the PQCD regulated scaling region. 
A quantitative definition of  duality is accomplished by comparing limited intervals (integrated in Bjorken-$x$ over the entire resonance region) defined according to the experimental data, {\it e.g.}~[\refcite{Niculescu:2000tk,Whitlow:1991uw}]. Hence, we compare the scaling results as a theoretical counterpart, or an output of PQCD, in the same kinematical intervals, and at the same scale $Q^2$. Namely we consider the ratio,
\begin{eqnarray}
\label{dual1}
R^{\mbox{\small Res/DIS}}(x_{\mbox{\tiny ave}}, Q^2)&=&\frac{
\int_{x_{\mbox {\tiny min}}}^{x_{\mbox {\tiny max}}} dx\,
F_2^{\mbox {\tiny data}} (x, Q^2)
}
{\int_{x_{\mbox{\tiny min}}}^{x_{\mbox{\tiny max}}} dx\,
F_2^{\mbox {\tiny DIS}} (x, Q^2)
}\quad.
\end{eqnarray}
We consider duality to be fulfilled if $R^{\mbox{\small Res/DIS}}(x_{\mbox{\tiny ave}}, Q^2)$ is $1$.
Note that the definition in Eq.(\ref{dual1}) relies on the fact that the PQCD evaluation, $F_2^{\mbox {\tiny DIS}} (x, Q^2)$, is very well constrained in the region of interest ($x \gtrsim 0.3$) despite it does not correspond directly to measured data. $F_2^{\mbox {\tiny DIS}}$ is an input that once fed into the evolution equations determines the structure functions behavior at much larger $Q^2$. However, the error on this type of backward evolution is expected here to be small, being dominated by the valence contribution (a quantitative analysis of the latter will be carried out in an upcoming study [\refcite{our}]).  Had we applied the same procedure to low $x$ where the singlet and gluon distributions govern $F_ 2$, we would have gotten a much larger error at low $Q^2$ because of the strong correlation with the value of $\alpha_s$. 

Besides perturbative evolution one has to take into account several hadronic corrections to $F_2^{\mbox {\tiny DIS}}$. For instance, if we evolve the structure functions to NLO, we find that duality is violated by a given amount. However Target Mass Corrections (TMCs) are important here and move the ratio closer to unity.
The most important effect for our purposes is the effect of LxR, that we develop hereafter.

\subsection{Large-$x$ Resummation}

Large $x$ threshold resummation
%%% LxR
effects (LxR) arise formally from terms containing powers of 
$\ln (1-z)$, $z$ being the longitudinal 
variable in the evolution equations, that are present in 
the Wilson coefficient functions $C(z)$. 
Below we write schematically how the latter relate the parton distributions to {\it e.g.} 
the structure function $F_2$, 
%%%
\begin{equation}
F_2^{LT}(x,Q^2)  = \frac{\alpha_s}{2\pi} \sum_q \int_x^1 dz \, C(z) \, q(x/z,Q^2), 
\label{lxr}
\end{equation}   
%%%
where we have considered only the non-singlet (NS) contribution to $F_2$ since 
only valence quarks distributions are relevant in our kinematics. 
The logarithmic terms in $C(z)$ become very large at large $x$, and they need to be 
resummed to all orders in $\alpha_s$. 
Resummation was first introduced by  
linking this issue to the definition of the correct kinematical variable that determines the 
phase space for the radiation of gluons
at large $x$. This was found to be $\widetilde{W}^2 = Q^2(1-z)/z$, 
instead of $Q^2$~[\refcite{Brodsky:1979gy,Amati:1980ch}].
As a result, the argument of the strong coupling constant becomes $z$-dependent: 
$\alpha_s(Q^2) \rightarrow \alpha_s(Q^2 (1-z)/z)$~[\refcite{Roberts:1990ww,Roberts:1999gb}].
In this procedure, however, an ambiguity is introduced, related to the need of continuing 
the value of $\alpha_s$  
for low values of its argument, {\it i.e.} for $z \rightarrow 1$~[\refcite{Pennington:1981cw}]. 

Since the size of this ambiguity is of the same order as the higher-twist corrections, it has been considered, in previous work~[\refcite{Niculescu:1999mw}], as a source of theoretical error or higher order effects. We propose  an accurate analysis~[\refcite{our}]  from which one can extract $\alpha_s$ for values of the scale in the infrared region. 
To do so, we investigate the effect of varying the form of the running coupling on the evolution equations. We consider the following choices:
\begin{itemlist}
\item $\alpha_s(Q^2)$ ;
\item an expansion of $\alpha_s(\tilde W^2)$ in $\ln((1-z)/z)$, to NLO,
	\begin{eqnarray}
		\alpha_s(\tilde W^2)=\alpha_s(Q^2)-\frac{\beta_0}{4\pi}\, \ln\left(\frac{1-z}{z}\right)\, \alpha_s^2(Q^2),
	\end{eqnarray}	
\item the complete $z$ dependence of $\alpha_s(\tilde W^2)$.
\end{itemlist}
%
%%%%%%%%%%%%%%%%%%%%%%%%%%%%%%%%%%%%%%%%%%%%%%%%%%%%%%%%
%          Fig.1 alphas with Q2
%%%%%%%%%%%%%%%%%%%%%%%%%%%%%%%%%%%%%%%%%%%%%%%%%%%%%%%%
\begin{figure}[pt]
\centerline{\psfig{file=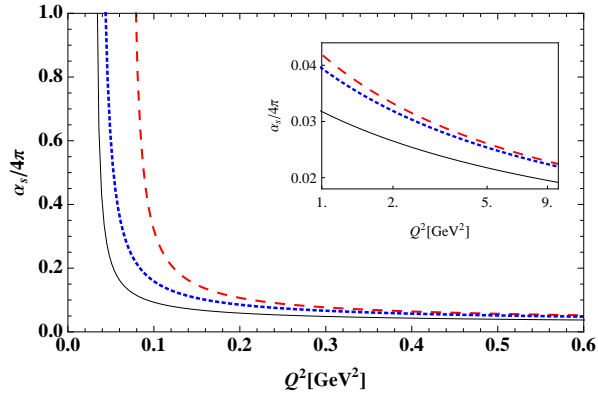,width=7.8cm}}
\caption{Running of the strong coupling constant in the $\overline{MS}$ with $\Lambda_{\mbox{\tiny LO}}=174$MeV. The solid black curve represent the LO $\alpha_s(Q^2)/4\pi$. The dashed red curve represents the expansion of the strong coupling in $\ln((1-z)/z)$, for $z=0.7$ ; the dotted blue curve is the complete $\alpha_s(Q^2(1-z)/z)/4\pi$ for the same value of $z$. }
\label{alphas_q}
\end{figure}%
%%%%%%%%%%%%%%%%%%%%%%%%%%%%%%%%%%%%%%%%%%%%%%%%%%%%%%%%
%

The running of each of the three versions of the coupling constant  starts being very different when $z\to 1$. We illustrate this behavior on Fig.~\ref{alphas_q} for $z=0.7$. The infrared behavior of the coupling constant with argument $Q^2$ starts to matter to lower $Q^2$ values than for a coupling constant which argument is $\tilde W^2$. On the other hand, the corresponding Landau poles do not coincide: for small $Q^2$ values and large-$z$, the argument of $\alpha_s(\tilde W^2)$ differs from the logarithmic terms taken into account in the NLO expansion. The asymptotic value differs for $\alpha_s(\tilde W^2)$ and expansion w.r.t. $\alpha_s(Q^2)$, as shown in the inner frame. 

\begin{figure}[pt]
\centerline{\psfig{file=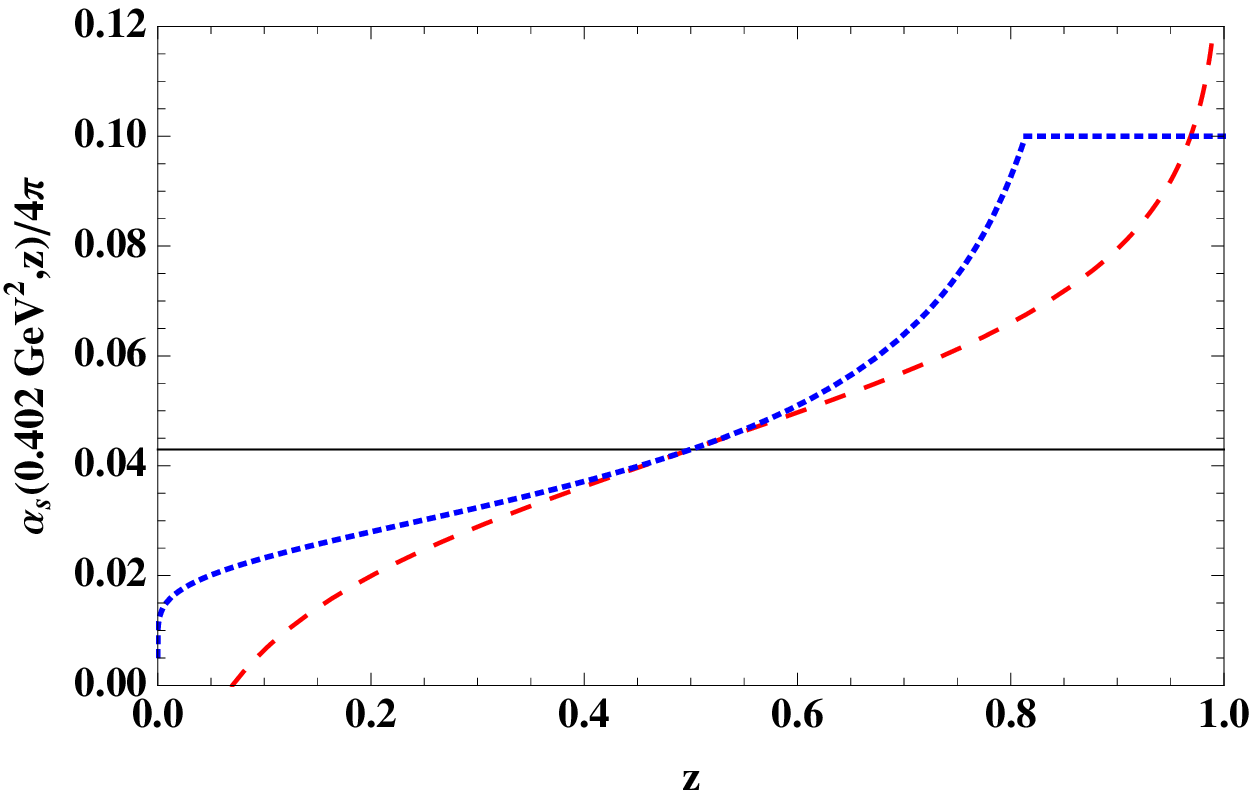,width=6cm}\psfig{file=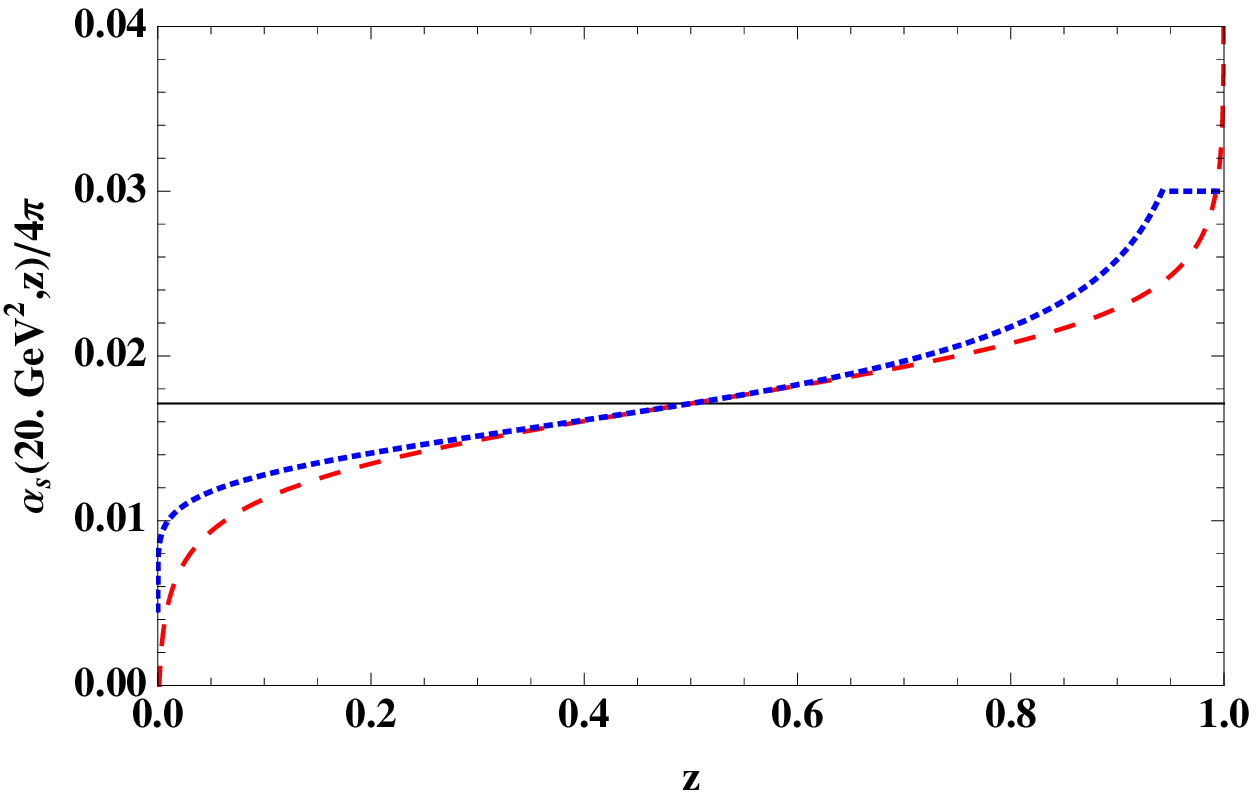,width=6cm}}
\caption{Running of the strong coupling constant in the $\overline{MS}$ with $\Lambda_{\mbox{\tiny LO}}=174$MeV at fixed $Q^2$. The solid black curve represent the LO $\alpha_s(Q^2)/4\pi$. The dashed red curve represents the expansion of the strong coupling in $\ln((1-z)/z)$ ; the dotted blue curve is the complete $\alpha_s(Q^2(1-z)/z)/4\pi$. The cut of the }
\label{alphas_z}
\end{figure}

The meaning of LxR becomes very clear from Fig.\ref{alphas_q}. It is now understood that the only free parameter in testing the realization of duality here, is related to $\alpha_s$. By playing with the argument of the running coupling constant, we can tune the scaling structure function and extract the low $Q^2$ behavior that determines duality. For instance, by setting a maximum value for $z$ one would prevent the DGLAP evolution from including extremely large values of the coupling constant. Moreover, this $z_{\mbox{\tiny max}}$ could define a criterion of convergence of the expansion w.r.t the complete $\alpha_s(\tilde W^2)$, as illustrated in Fig.~\ref{alphas_z}.

\begin{figure}[pb]
\centerline{\psfig{file=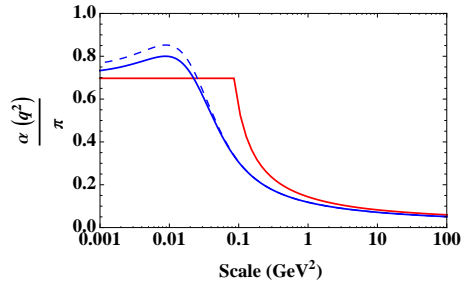,width=6.cm}}
\caption{The running coupling constant. The solid red curve represents the  perturbative coupling constant in $\overline{MS}$ with $\Lambda_{\mbox{\tiny LO}}=174$ MeV. The blue curve represent the Cornwall effective charge of Eq.~(\ref{alphalog}) for the same value of $\Lambda$, $m_0^2=1.05 \Lambda^2$ and for $\rho=1.7,\, 1.66$ on, respectively, the solid and dashed curves.
}
\label{compar}
\end{figure}

This exercice has to be repeated for each experimental data point. We observe from our analysis that the maximum value of $z$ or, equivalently, the scale in which the running of the coupling is stopped  changes from one to another data point. A rough qualitative parameterization of the realization fo duality would look like Fig.~\ref{compar}, where we have use the Cornwall's effective charge resulting from a massive gluon propagator~[\refcite{Cornwall:1982zr}]. In effect, the dynamical gluon mass generation leads  to the  freezing of the QCD running coupling constant. The  nonperturbative  generalization  of $\alpha_s(Q^2)$ comes, here, in the form
%%%%%%%%%%%%%%%%%%%%%%%%%%%%%%%%%%%%%%%%%%%%%%%%%%%%%%%%
%          Fig.2 alphas with Q2
%%%%%%%%%%%%%%%%%%%%%%%%%%%%%%%%%%%%%%%%%%%%%%%%%%%%%%%%
%%%%%%%%%%%%%%%%%%%%%%%%%%%%%%%%%%%%%%%%%%%%%%%%%%%%%%%%
%
%%%%%%%%%%%%%%%%%%%%%%%%%%%%%%%%%%%%%%%%%%%%%%%%%%%%%%%%
%          Fig.3 cornwall alpha
%%%%%%%%%%%%%%%%%%%%%%%%%%%%%%%%%%%%%%%%%%%%%%%%%%%%%%%%
%%%%%%%%%%%%%%%%%%%%%%%%%%%%%%%%%%%%%%%%%%%%%%%%%%%%%%%%
\vspace{-.1cm}
\begin{equation}
\frac{\alpha_{\mbox{\tiny NP}}(Q^2)}{4\pi} = \left[\beta_0 \ln \left(\frac{Q^2 +\rho m^2(Q^2)}{\Lambda^2}\right)\right]^{-1} \quad.
\label{alphalog}
\end{equation}
The zero gluon mass limit leads to the LO perturbative coupling constant momentum dependence.
The $m^2(Q^2)$ in the argument of the logarithm 
\begin{equation}
m^2 (Q^2)= m^2_0\left[\ln\left(\frac{Q^2 + \rho m_0^2}{\Lambda^2}\right)
\bigg/\ln\left(\frac{\rho m_0^2}{\Lambda^2}\right)\right]^{-1 -\gamma},
\label{rmass}
\end{equation}
with ${(\gamma)} = 1/11$, tames  the  Landau pole. The mass
$m^2(Q^2)$ can be understood as a constituent gluon mass, and depends non-trivially  on the momentum  transfer $Q^2$.
As a consequence $\alpha_s(Q^2)$ freezes 
at a  finite value in the IR~[\refcite{Cornwall:1982zr,Aguilar:2009nf}].%, namely  
%\mbox{$\alpha_s^{-1}(0)/4\pi= \beta_0 \ln (\rho m^2(0)/\Lambda^2)$}~
%\\

An analysis of the complete data set collected at JLab in addition to the existing large $x$ data will allow us to carry our study to a quantitative level~[\refcite{our}]. Three nonperturbative representations of the coupling constant will there be analyzed~[\refcite{Cornwall:1982zr}--\refcite{Shirkov:1997wi}].

\section{The other way: From nonperturbative to perturbative QCD}

Now, trying to match nonperturbative model with PQCD evolution,  we examine the consequences on the hadronic scale in considering the same Cornwall's effective charge in the evolution of the PDFs, evaluated in hadronic models.

In QCD all matrix elements must have a scale associated to them as a result of the RGE  of the theory.  A fundamental step in the development of the use of hadron models for the description of  properties at high momentum scales was the assertion that all calculations done in a model should have a  RGE scale associated to it~[\refcite{Jaffe:1980ti}]. The momentum distribution inside the hadron is only related to the dynamical scale and not to the momentum governing the RGE. Thus a model calculation only gives a boundary condition for the RG evolution as can be seen for example in the LO evolution equation for the moments of the valence quark distribution 
\begin{equation}
\langle q_v(Q^2)\rangle_n = \langle q_v(\mu_0^2)\rangle_n \left(\frac{\alpha_s{(Q^2)}}{\alpha_s{(\mu_0^2)}}\right)^{d^n_{NS}},
\label{moments}
\end{equation}
where $d^n_{NS}$ are the anomalous dimensions of the NS distributions.  For calculations in the bag model, the dynamics  it describes is unaffected by the evolution procedure, and the model provides only the expectation value,  $\langle q_v(\mu_0^2)\rangle_n$, which is associated with the hadronic scale. 
When considering the nonperturbative solution of the Dyson--Schwinger equations, that results in the appearence of an infrared cut-off,  the gluon mass Eq.~(\ref{rmass}) will only affect the evolution of the PDFs. The generalization of the coupling constant results to the structure function imply that the LO evolution Eq.~(\ref{moments})  simply changes by incorporating  the nonperturbative coupling constant  evolution Eq.~(\ref{alphalog}). In other words, the hadronic scale that ensues from the nonperturbative procedure is quantitatively different  from the perturbative scheme. However, the results from both procedure are close, what give us confidence on the perturbative procedure even at low scales.
We note however, that the corresponding hadronic scale, for the sets of parameters chosen in Fig.~\ref{compar}, turns out to be slightly smaller than in the perturbative case ($\mu_0^2 \sim 0.1$ GeV$^ 2$).   The physical meaning here would be that the nonperturbative approach  seems to favor a scenario where at the hadronic scale we have not only valence quarks but also gluons and sea quarks. 

The application of the nonperturbative framework  to the evaluation of the T-odd TMDs illustrates the uncertainty on model predictions coming only from the matching with RGE. In Ref.~[\refcite{Courtoy:2011mf}] we have considered the errorband resulting from the uncertainty of the initial value of  $\alpha_s(\mu_0)$ on the bag model evaluation of the Sivers and Boer-Mulders functions. 
This observation shows that the naive scenario may well serve to make predictions, within a reasonably small band, which should not be far from experimental expectations. Within this interpretation, T-odd TMDs are possible candidates to study the behavior of $\alpha_s$ at intermediate and low $Q^2$, though still biased by strong theoretical difficulties such as the TMD evolution~[\refcite{Aybat:2011ge}] or ambiguities due to the choice of the hadronic representation.

\section{Conclusions}

 A combined analysis of the extractions of the running coupling constant in the infrared region suggests a novel definition of the effective charge~[\refcite{our}], following the example of Ref.~[\refcite{Deur:2005cf}]  where the effective coupling constants are phenomenologically inferred from different processes, and to calculations based on Schwinger--Dyson equations. This analysis also bears potential important consequences for the  connection of low scale/hadronic models with experiments in the multi-GeV region. 
%

%%%%%%%%%%%%%%%%%%%%%%%%%%%%%%%%%%
\section*{Acknowledgments}

We are grateful to J.P. Chen, A. Deur, and V. Vento  for fruitful discussions.

%\begin{thebibliography}{000} %for 3 digits
%\begin{thebibliography}{00}  %for 2 digits

\end{document}